\documentclass[fleqn,usenatbib]{mnras}

\usepackage{newtxtext,newtxmath}

\usepackage[T1]{fontenc}

\DeclareRobustCommand{\VAN}[3]{#2}
\let\VANthebibliography\thebibliography
\def\thebibliography{\DeclareRobustCommand{\VAN}[3]{##3}\VANthebibliography}

\usepackage{graphicx}	
\usepackage{amsmath}	
\usepackage{multirow} 
\usepackage[normalem]{ulem} 

\newcommand\Eqn[1]{Eq.~(\ref{#1})} 

\newcommand{\nn}{\nonumber}
\newcommand{\be}{\begin{equation}}
\newcommand{\ee}{\end{equation}}
\newcommand{\beq}{\begin{eqnarray}}
\newcommand{\eeq}{\end{eqnarray}}

\newcommand{\ep}{\varepsilon}

\newcommand{\bv}{Brunt--V\"ais\"al\"a}
\newcommand{\gm}{$g$-mode}
\newcommand{\gms}{$g$-modes}
\newcommand{\cs}{c_s}
\newcommand{\ce}{c_e}

\title[DM $g$-modes]{$g$-mode oscillations of dark matter admixed neutron stars}

\author[Shirke et al.]{
Swarnim Shirke,$^{1}$\thanks{E-mail: swarnim@iucaa.in}
Debarati Chatterjee,$^{1}$
and Prashanth Jaikumar$^{2}$
\\
$^{1}$Inter-University Centre for Astronomy and Astrophysics, Post Bag 4, Ganeshkhind, Pune University Campus,
Pune - 411007, India\\
$^{2}$Department of Physics and Astronomy,
California State University Long Beach, Long Beach, California 90840 U.S.A.\\
}

\date{Accepted XXX. Received YYY; in original form ZZZ}

\pubyear{2025}

\begin{document}
\label{firstpage}
\pagerange{\pageref{firstpage}--\pageref{lastpage}}
\maketitle

\begin{abstract}
We investigate \gm~oscillations in dark matter admixed neutron stars employing a relativistic mean field model to describe hadronic matter and a model for self-interacting fermionic dark matter motivated by the neutron decay anomaly. Following the construction of such admixed configurations, we derive the equilibrium and adiabatic speeds of sound therein, leading to a computation of the star's \gm~spectrum in the Cowling approximation. In particular, we explore the effect of dark matter self-interaction, the nucleon effective mass and dark matter fraction on the principal \gm~frequency, and its first overtone. We show that the effect on \gm~frequency depends predominantly on the dark matter fraction, and demonstrate an equation of state-independent constraint for the latter. Prospects of identifying the presence of dark matter in neutron stars using \gms~are discussed.
\end{abstract}

\begin{keywords}
Asteroseismology -- Neutron Stars -- Dark Matter -- Gravitational Waves
\end{keywords}



\section{Introduction}

The origin and nature of dark matter (DM) is one of the fundamental and unresolved problems in physics~\citep{Cirelli2024}. While indirect evidence for DM comes mainly from the observation of galaxy rotation curves, galaxy clusters, large-scale structure and the cosmic microwave background, the study of compact astrophysical objects as a probe for its detection has gained interest over the years~\citep{RajTanedoYu2018,Baryakhtar2022, Bramante2024, LeaneTong2024}. DM can accumulate around compact objects like black holes and neutron stars (NSs), getting trapped in their deep gravitational potential, forming a DM core in their interior or a halo around them. Accumulation of DM on ordinary stars and planets has also been explored~\citep{ellis2018, Horowitz2019ns, Horowitz2020sun, Horowitz2020earth}.

NSs, as compact remnants of Type II supernovae with interior densities of the order of nuclear saturation density, provide a natural environment to study properties of dense matter beyond the scope of terrestrial nuclear or heavy-ion laboratories. A theoretical study of DM admixed NSs is facilitated by a model of cold and dense neutron-rich matter beyond laboratory conditions~\citep{GlendenningBook}. Here, we adopt a phenomenological mean-field approach to determine the equation of state (EoS) (pressure-density relationship) of the NS interior~\citep{GlendenningBook,Schaffner-Bielich_Book}. Global properties of the NS, such as its mass or radius, follow from a solution of the hydrostatic equilibrium (TOV) equations, which can then be compared to astrophysical data to impose constraints on the model parameters. From multi-wavelength astrophysical observations of NSs in binary systems, it is possible to determine the maximum NS mass, a constraint that should be satisfied for any given EoS model~\citep{GRMHDBook}. While NS masses can be determined to good accuracy in binary systems from the measurement of relativistic effects, the determination of NS radius is particularly challenging. Recent progress in the determination of radii has been achieved by the NICER mission employing the pulse profile modelling technique~\citep{GRMHDBook}.

Of particular relevance to this work is the fact that gravitational waves (GWs) emitted by NSs offer the possibility of directly probing their interior. GW detection from the binary NS merger event GW170817~\citep{Abbott2017AGW170817}, along with its electromagnetic counterparts~\citep{abbott2017BGW170817multi}, provided an independent constraint on the NS radius, through the measurement of the tidal deformation of the merger components during the inspiral phase~\citep{Abbott2018}. However, no other such multi-messenger candidates have been conclusively observed yet by the LIGO-Virgo-KAGRA collaboration. In addition to mergers, another way to study these compact objects using GWs is by studying the emission from quasi-normal oscillation modes in isolated or binary NSs~\citep{Andersson_GW_Book}. These oscillations are classified into various fluid modes e.g. {\it f-,p-,g-} and {\it r-} modes, based on the restoring forces involved. Stellar properties can be inferred by analysing these oscillations using asteroseismology.

The effects of admixture of DM in NSs have been studied in various contexts including asteroseismology, particularly for $f$-mode oscillations~\citep{Das2021, Shirke2024, Flores2024, Sen2024, Jyothilakshmi2024, Dey2024, Thakur2024, Husain2025, Sotani2025, Kumar2025}. In a recent work, we explored the effect of DM admixed NSs on $r$-modes~\citep{Shirke2023b}. However, to our knowledge, there is no study exploring the \gms~for DM admixed NSs, so we have taken up the task in this paper. 
The primary mechanism for an admixture of DM with NSs is accretion, but it is known that this process can only accumulate a small amount of DM $\sim10^{-10}M_\odot$ over Hubble time~\citep{ellis2018}.
Other mechanisms resulting in larger DM fractions have been proposed recently, e.g., via the decay of neutrons~\citep{FornalGrinstein2018prl}. This is motivated by the neutron decay anomaly observed in experimental particle physics, which refers to the discrepancy in the decay time of neutrons when observed using two different methods: i) bottle experiments that count the number of neutrons and ii) beam method that counts the number of protons appearing via $\beta$ decay ($n \rightarrow p + e^- + \bar{\nu}_e$). A possible resolution to the problem is that some fraction of the neutrons decay via the dark sector~\citep{FornalGrinstein2018prl}. This would have important consequences on NS physics, as NS cores are predominantly composed of unbound neutrons, leading to the formation of DM admixed NSs~\citep{Berryman2022, Shirke2023b, GardnerZakeri2023}. This scenario can lead to large DM fractions in NSs and has been used to study the properties of DM admixed NSs~\citep{motta2018a, motta2018b, Husain2022a}. 
In a recent publication~\citep{Shirke2023b}, some authors of this work employed this model to explore the relativistic instability induced via the CFS mechanism and proposed rotational $r$-mode oscillations as a novel probe for DM in NSs. The non-radial fundamental $f$-mode oscillation, which couples strongly to gravitational radiation, as well as dynamical tidal fields, was explored in~\cite{Shirke2024}. In the present work, we extend these studies to investigate the effect of DM admixture on $g$-modes in NSs.
 
As a means to investigate a potentially observable signature of DM in NSs, we look to \gm~oscillations as they are particularly sensitive to composition, and changes thereof. 
A \gm~is a specific type of fluid oscillation where a parcel of fluid is displaced against the background of a stratified environment inside a NS. While mechanical equilibrium is rapidly restored, chemical equilibrium can take longer, causing buoyancy forces to oppose the displacement. Since cold NSs are not convective, the opposing force sets up stable oscillations with a typical frequency called the (local) \bv~frequency. The kind of core \gms~we study here were introduced in~\cite{Reisenegger92,Reisenegger94,Lai1994,Lai1999} and, in a recent work,~\cite{Jaikumar:2021jbw} showed that the \gm~frequency rises steeply with the onset of quarks due to a rapid change in the equilibrium sound speed (see also~\cite{wei2020lifting}). Since \gm~oscillations couple to tidal forces, they may be excited during the merger of two NSs and provide information on the interior composition, specifically DM admixed with ordinary matter. Simulations are also being carried out to probe the effect of DM admixture on the evolution of NSs in binary systems~\citep{Emma2022, Bauswein2023, Hannes2023}. Details about the calculation of the \gm~are given in the next section.

For a chosen equation of state that admits DM, we compute the global solutions representing the discrete $g$-mode spectrum of the NS and examine trends in the fundamental \gm~frequency as a function of DM coupling. As in other works on the \gm~\citep{Constantinou2021, Jaikumar:2021jbw, Kantor:2014lja, Reisenegger92, McDermott83}, we use the Cowling approximation~\citep{Cowling-Approx}, which neglects the back reaction of the gravitational potential but includes relativistic effects of matter~\citep{Kantor:2014lja}. We note that the impact of the Cowling approximation, compared to a full general relativistic (GR) calculation, typically only affects the frequencies of the \gm\, at the 5-10\% level~\citep{Grig} for stars around 2-2.25$M_{\odot}$~\citep{Zhao2022} (also see appendix~\ref{sec:appendix_fullGR}), therefore it will not change our conclusions qualitatively. This is much lower than the $30\%$ deviation observed in the case of pressure $f$-modes~\citep{Pradhan2022}. The global \gm~frequency for a given stellar configuration (fixed gravitational mass $M$ and radius $R$) can be thought of as an average of the local \gms~(although it is still sensitive to phase transitions).

The model considered involves neutron dark decay, which has timescales in hours~\citep{FornalGrinstein2018prl} for free neutrons. This could vary depending on the nuclear interaction involved~\citep{McKeen2018}. 
This timescale is much lower than the typical age of NSs, leading to formation of equilibrated DM admixed NSs within their lifetimes that characterizes the structure of the NS (discussed in Sec.~\ref{sec:formalism}). On the other hand, it is larger than typical \gm~oscillation time periods with frequencies of $\mathcal{O}(100)$ Hz driving the system out of chemical equilibrium during oscillations. This disturbance in the chemical equilibrium is essential to drive the \gms~, however, it does not alter the overall NS chemical equilibrium and structure.

In this work, we study the fundamental $g$-mode with radial quantum number $n$ = 1 and fix the mode's multipolarity at $l$ = 2. This is because the $l$ = 2 mode is quadrupolar in nature, and can couple to GWs. Higher $l$ values (octupole and higher) are generally weaker than the quadrupole. 
The reason to study the $n$ = 1 (fundamental) $g$-mode is that the local dispersion relation for $g$-modes $\omega^2\propto 1/k^2$ implies that the $n$ = 1 excitation has the highest frequency, whereas higher values of $n$ are known to have a smaller amplitude of excitation and a weaker tidal coupling coefficient~\citep{Constantinou2021}. The fundamental $g$-mode is also within the sensitivity range of the current generation of GW detectors~\citep{Lai1994,ZhaoLattimer2022}.

The paper is structured as follows: The details of the models describing nuclear matter, DM, and their parameters, along with the computational details of \gms~and the speed of sound calculations, are provided in Sec.~\ref{sec:formalism}. We report and discuss the results of this work in Sec.~\ref{sec:results}, summarizing the conclusions in Sec.~\ref{sec:conclusions}.

\section{Formalism}
\label{sec:formalism}

Here, we describe the particular models used to describe the hadronic matter and DM in NSs for the study of $g$-modes. We then discuss the choice of model parameters and then provide the details of the formalism used to calculate observables and $g$-modes.

\subsection{Model for Hadronic Matter}
Hadronic matter is described using the phenomenological Relativistic Mean-Field (RMF) model, in which the strong interaction between the nucleons ($N$), i.e., neutrons ($n$) and protons ($p$), is mediated via exchange of scalar ($\sigma$), vector ($\omega$) and iso-vector ($\boldsymbol{\rho}$) mesons. The corresponding interaction Lagrangian is~\citep{hornick2018}
\begin{align}\label{interaction_lagrangian}
\mathcal{L}_{int} &= \sum_{N} \bar\psi_{N}\left[g_{\sigma}\sigma-g_{\omega}\gamma^{\mu}\omega_{\mu}-\frac{g_{\rho}}{2}\gamma^{\mu}\boldsymbol{\tau\cdot\rho}_{\mu}\right]\psi_{N} \nonumber \\
&-\frac{1}{3}bm(g_{\sigma}\sigma)^{3}-\frac{1}{4}c(g_{\sigma}\sigma)^{4} \nonumber \\ 
&+ \Lambda_{\omega}(g^{2}_{\rho}\boldsymbol{\rho^{\mu}\cdot\rho_{\mu}})(g^{2}_{\omega}\omega^{\nu}\omega_{\nu}) + \frac{\zeta}{4!}(g^{2}_{\omega}\omega^{\mu}\omega_{\mu})^{2}~,
\end{align}
where $\psi_N$ is the Dirac spinor for the nucleons, $m$ is the vacuum nucleon mass, $\{\gamma^{i}\}$ are the gamma matrices, $\boldsymbol{\tau}$ are Pauli matrices, and $g_{\sigma}$, $g_{\omega}$, $g_{\rho}$ are meson-nucleon coupling constants.  $b$, $c$, and $\zeta$ are the scalar and vector self-interactions couplings respectively, and $\Lambda_{\omega}$ is the vector-isovector interaction. $\zeta$ is set to zero as it is known to soften the EoS~\citep{mueller1996, tolos2017, pradhan2022zeta}. The energy density for this RMF model is given by~\citep{hornick2018}
\begin{small}
\begin{align}\label{energydensity}
\epsilon_{nuc}&=\sum_{N}\frac{1}{8\pi^{2}}\left[k_{F_{N}}E^{*3}_{F_{N}}+k^{3}_{F_{N}}E^*_{F_{N}}
-m^{*4}\ln\left(\frac{k_{F_{N}}+E^*_{F_{N}}}{m^{*}}\right)\right] \nonumber \\
&+\frac{1}{2}m^{2}_{\sigma}\bar\sigma^{2}+ \frac{1}{2}m^{2}_{\omega}\bar\omega^{2}+\frac{1}{2}m^{2}_{\rho}\bar\rho^{2} \nonumber \\ 
&+ \frac{1}{3}bm(g_{\sigma}\bar\sigma)^{3} + \frac{1}{4}c(g_{\sigma}\bar\sigma)^{4} + 3\Lambda_{\omega}(g_{\rho}g_{\omega}\bar\rho\bar\omega)^{2} + \frac{\zeta}{8}(g_{\omega}\bar{\omega})^{4}~,  
\end{align}
\end{small}where $k_{F_{N}}$ is the Fermi momentum, $E^*_{F_{N}} = \sqrt{k_{F_{N}}^{2} + m^{*2}}$ is the Fermi energy, and $m^{*}=m-g_{\sigma}\sigma$ is the effective mass. All mesonic fields are replaced by their mean values. The following equations of motion are used to solve for these mean field values,
\begin{align} \label{mesonic_eoms}
   m_\sigma^2  \bar \sigma + m \, b \,  g_{\sigma}^3 \bar \sigma^2 + c \,
     g_{\sigma}^4 \bar \sigma^3 &=  g_{\sigma} \, n^s , \nonumber \\  
   m_\omega^2 \, \bar \omega + \frac{\zeta}{3!}  g_{\omega}^4 \bar \omega^3
     + 2 \Lambda_{\omega} g_{\rho}^2  g_{\omega}^2  \bar \rho^2 \bar \omega &=
     g_{\omega} \, n , \nonumber \\  
   m_\rho^2 \,  \bar \rho + 2 \Lambda_{\omega} g_{\rho}^2  g_{\omega}^2
     \bar \omega^2 \bar \rho &=   \frac{g_{\rho}}{2} \, n_3 \ .  
\end{align}
Here, $n^s=n^s_p+n^s_n$, $n=n_p+n_n$, and $n_3=n_p-n_n$. The densities $n^s_{i}$ and $n_i$ are defined as
\begin{align}
n^s_{i}&=\frac{m^{*}}{2 \pi^{2}} \left[E^*_{F_i}k_{F_i}-{m}^{*2} \ln
           \frac{k_{F_i}+E^*_{F_i}}{m^{*}} \right] \ , \nonumber \\ 
n_i&=\frac{k_{F_i}^{3}}{3\pi^{2}} ,
\end{align} 

 The matter is in weak beta equilibrium and charge neutral, resulting in the following conditions,
\begin{equation}\label{eqn:betaeqlbm_neutrality}
    \mu_n = \mu_p + \mu_e,~ \mu_\mu = \mu_e,~ n_p = n_e + n_\mu~.
\end{equation}

\subsection{Model for Dark Matter}
For DM, we employ a model motivated by the neutron decay anomaly.~\cite{FornalGrinstein2018prl} suggested that the anomaly could be explained if about $1\%$ of the neutrons decayed to DM. Multiple decay channels were proposed, e.g. $n \rightarrow \chi + \phi$, $n \rightarrow \chi + \chi + \chi$, $n \rightarrow \chi + \gamma$~\citep{FornalGrinstein2018prl, Strumia2022}. We consider the most widely studied one, where the neutron decays into a dark fermion with baryon number one and a light dark boson:
\begin{equation}
    n \rightarrow \chi + \phi
\end{equation}
Alternative decay channels such as $n \rightarrow \chi + \chi + \chi$ could also be considered~\citep{Strumia2022}. The light dark particle $\phi$ with its mass $m_{\phi}$ set to zero escapes the NS, and chemical equilibrium is established via $\mu_N = \mu_{\chi}$. Various stability conditions require the mass of the DM particle ($m_{\chi}$) to be in a narrow range of $937.993 < m_{\chi} < 938.783$~\citep{Shirke2023b}. We set $m_{\chi} =  938.0$ MeV. We further add self-interaction $G$ between DM particles mediated via a vector gauge field $V_{\mu}$.
 The energy density of DM is given by
\begin{equation}
    \epsilon_{\chi} = \frac{1}{\pi^2}\int_{0}^{k_{F_{\chi}}} k^2\sqrt{k^2 + m_{\chi}^2}dk + \frac{1}{2}Gn_{\chi}^2~,
    \label{eqn:endens_dm}
\end{equation}
where,
\begin{equation} \label{eqn:Gdefinition}
    G = \left(\frac{g_V}{m_V}\right)^2, \qquad n_{\chi} = \frac{k_{F_{\chi}}^3}{3 \pi^2}~.
\end{equation}
Here, $g_V$ is the coupling strength, $m_V$ is the mass of the vector boson. $k_{F_{\chi}}$ and $n_{\chi}$ are the Fermi momentum and number density of the DM particles, respectively. From this, we obtain the chemical potential
$\mu_{\chi} = \sqrt{k_{F_{\chi}}^2 + m_{\chi}^2} + Gn_{\chi}$. We add this contribution ($\epsilon_{\chi}$) to the energy density of hadronic matter ($\epsilon_{nuc}$) to get the total energy density ($\epsilon=\epsilon_{nuc}+\epsilon_{\chi}$) and calculate the pressure using the Gibbs-Duhem relation
\begin{equation} \label{eqn:pressure}
P = \sum_{N}{}\mu_{N}n_{N} + \mu_\chi n_\chi- \epsilon~,
\end{equation}
where, $\mu_{N} = E_{F_{N}} + g_{\omega}\bar{\omega} + \frac{g_{\rho}}{2}\tau_{3N}\bar{\rho}$. We further have free fermionic contributions from the leptons ($l$), i.e., electrons ($e$) and muons $(\mu)$ to the energy density and pressure. We vary the baryon density ($n_b = n_p + n_n + n_{\chi}$) and compute the EoS subject to the conditions in Eqs.~(\ref{eqn:betaeqlbm_neutrality}) and $\mu_N =\mu_{\chi}$.

We define the DM fraction  ($f_{DM}$) as 
\begin{equation}\label{eqn:dm_frac}
    f_{DM} = \frac{\int_V \epsilon_{\chi} dV}{\int_V \epsilon dV}~,
\end{equation}
where the integration is carried out over the entire star's volume. Note that DM fraction mostly depends on the stellar mass and the interaction parameter $G$~\citep{Shirke2024}.

In our work, we consider neutron dark decay only in the core. The assumed DM particle mass prohibits the decay of neutrons within nuclei~\citep{Fornal2020}. Hence, we do not consider any dark component in the crust. 
We neglect the neutron decay in the inner crust as the abundance of free neutrons is very low. 
We use the EoS from~\cite{Hempel2010} for the crust, which is smoothly connected to the core EoS, ensuring causality and thermodynamic consistency.

\subsection{Model Parameters}
We have a total of eight coupling parameters in this model, six from the hadronic model ($g_{\sigma}$, $g_{\omega}$, $g_{\rho}$, $b$, $c$, $\Lambda_{\omega}$) and two ($g_V$, $m_V$) from the DM model. However, not all of them are independent. In the EoS of DM, the additional contribution due to the vector self-interaction is proportional to the parameter $G = g_V/m_V$.
Thus, we effectively only have one parameter in the dark sector: $G$. So, in all, there are seven independent model parameters, six controlling the hadronic part and one controlling the dark sector.

\begin{table*}
    \centering

    \begin{tabular}{|c|c|c|c|c|c|c|}
    \hline
       Model& $n_0$ $\rm (fm^{-3})$& $E_{sat}$ (MeV) & $K_{sat}$  (MeV) & $J$ (MeV) & $L$ (MeV) & $m^*/m$ \\ \hline \hline
        RMF~\citep{hornick2018} & 0.15 & -16.0 & 240 & 31  & 50&[0.55, 0.60, 0.65, 0.70, 0.75]\\
        \hline
       Stiffest~\citep{Ghosh2022EPJA} & 0.145 & -15.966 & 238.074 & 31.080 & 56.48& 0.550\\
       \hline
    \end{tabular} 
      \caption{\label{table:parameters}%
    Range of the variation of the nuclearsa parameters used in this work. The values of the DM self-interaction parameter $G$ corresponding to each of the nuclear EoS parameter sets is provided in Table~\ref{table:DM_parameters}.}
\end{table*}

The six coupling parameters of the hadronic model are fitted to nuclear saturation parameters. It was shown in previous works for RMF EoSs that among the nuclear parameters, it is the effective nucleon mass $m^*/m$ which is the dominant one that controls the NS observables, including oscillation mode characteristics~\citep{Ghosh2022EPJA,Pradhan2021}.
The ``RMF" set in the table, therefore, denotes the median values of the uncertainty ranges in nuclear saturation parameters considered in a previous work~\citep{hornick2018}, while only the effective nucleon mass is varied within its uncertainty range. The ``Stiffest" set in the table corresponds to the stiffest parameter set in the posteriors resulting from a Bayesian analysis in~\cite{Ghosh2022EPJA}, which satisfies the chiral effective field theory (CEFT) as well as astrophysical data. 
In all these models, $\Lambda_\omega$ has been set to zero. The values of the nuclear parameters in this work have been tabulated in Table.~\ref{table:parameters}.

The effect of the presence of DM on the NS EoS has been systematically investigated in our previous works~\citep{Shirke2023b,Shirke2024}. The role of the self-interaction parameter $G$ on the EoS and static observable properties of NSs can be understood from Figs.~1-7 of~\cite{Shirke2023b} or Figs.~1-2 of~\cite{Shirke2024}. Similarly, the effect of DM on $f$-modes can be seen in Figs.~3-7 of ~\cite{Shirke2024}. It was concluded from these studies that the effective mass $m^*/m$ is the most dominant nuclear parameter to dictate the macroscopic properties, while among the DM parameters, we found a strong
correlation only between the DM fraction and G.
Therefore, to investigate the effect of DM on $g$-modes, for each hadronic EoS, we consider discrete values of $G$ within their allowed range, i.e., for which the NS maximum mass exceeds $2M_{\odot}$. For example, for the ``Stiffest" parametrization, the values used are $G=6,15,30,100$ and $300$ fm$^2$. For the ``RMF" set, refer to Table~\ref{table:DM_parameters} for the values used for each $m^*/m$. 
The lower cutoff in each case is set by the two solar mass constraint. For each model, we also consider the case without any DM (no/without DM), which also corresponds to the case $G \to \infty$.
\begin{table}
    \centering
    \begin{tabular}{|c|c|c|}
    \hline
       Nuclear EoS & $m^*/m$&$G$ [fm$^2$] \\ \hline \hline
        Stiffest & - & [6,15,30,100,300]\\
        \hline
       \multirow{5}{2em}{RMF} & $0.55$ & [6,10,15,30,100,300]\\
       & $0.60$ & [7,10,15,30,100,300]\\
       & $0.65$ & [9,15,30,100,300]\\
       & $0.70$ & [14,20,30,50,100,300]\\
       & $0.75$ & [36,50,100,300]\\
       \hline
    \end{tabular} 
      \caption{\label{table:DM_parameters}%
    Values of $G$ used for various nuclear EoS parameters considered in this work.}
\end{table}

\subsection{Metric Functions}
Since DM particles are in chemical equilibrium with neutrons, and the DM density profiles follow the hadronic one, we assume a single fluid system.
For a spherically symmetric non-rotating NS, the line element is given by $$ds^2 = -e^{-\nu(r)}dt^2 + e^{\lambda(r)}dr^2 + r^2d\Omega^2~.$$ The macroscopic properties and the profiles of the metric functions are obtained by solving the Tolman-Oppenheimer-Volkoff (TOV) equations~\citep{Tolman1939, OppenheimerVolkoff1939}
\begin{align}\label{eqn:TOV}
    \frac{dm}{dr} &= 4\pi r^2 \epsilon(r) \nonumber \\
    \frac{dp}{dr} & = -\frac{(p(r)+\epsilon(r))}{2}\frac{d\nu}{dr} \nonumber \\
    \frac{d\nu}{dr} &= 2\frac{m(r)+4\pi r^3p(r)}{r(r-2m(r))} 
\end{align}

\subsection{Calculation of $g$-modes}

The \gm~frequencies ($\nu_g$=$\omega/(2\pi)$) are estimated within the relativistic Cowling approximation by computing numerical solutions to the following equations of motion for fluid variables $U,V$~\citep{Jaikumar:2021jbw}
\beq
\label{eq:uv}
\frac{dU}{dr}&=&\frac{g}{c_s^2}U+{\rm e}^{\lambda/2}\left[\frac{l(l+1){\rm e}^{\nu}}{\omega^2}-\frac{r^2}{c_s^2}\right]V \nn \\ 
\frac{dV}{dr}&=&{\rm e}^{\lambda/2-\nu}\,\frac{\omega^2-N^2}{r^2}U+g\Delta(c^{-2})V \,,
\eeq
which are simplified forms of the original fluid perturbation equations~\citep{McDermott83,Reisenegger92,Kantor:2014lja}. 
In \Eqn{eq:uv}, $U$ = $r^2{\rm e}^{\lambda/2}\,\xi_r$, $V$ = $\omega^2 r\, \xi_h$ = $\delta P/(\ep+P)$,   $\Delta(c^{-2})=\ce^{-2}-\cs^{-2}$, where $\xi_r$, $\xi_h$ are radial and tangential components of the fluid displacement,  $\delta P$ is the Eulerian pressure perturbation. The scale of the mode frequency is set by the \bv~frequency 
\beq \label{eqn:brunt_vaisala_frequency}
N^2 = g^{2}\Delta(c^{-2}){\rm e}^{\nu-\lambda},
\eeq
where the local gravitational field $g=-\nabla P/(\ep+P)$. 

The relativistic Cowling approximation neglects the back reaction of the gravitational potential by excluding metric perturbations that must accompany matter perturbations in a GR treatment~\citep{Thorne:1967a,Thorne:1967b,Lindblom:1983,Detweiler:1985,Finn:1987,Andersson:1995wu}. 
It reduces the number and complexity of the equations we have to solve while providing results for \gm~frequencies that are accurate at the few percent level~\citep{Grig} (also see Appendix~\ref{sec:appendix_fullGR}).
Details on the solution methods for \Eqn{eq:uv} and relevant boundary conditions are provided in~\cite{Jaikumar:2021jbw}.

\subsection{Adiabatic sound speed calculation}
The \gm~spectrum depends on two speeds of sound, the equilibrium and the adiabatic, where particle fractions are held fixed. The equilibrium speed of sound ($c_e$) is straightforward to calculate. This is obtained directly from the beta-equilibrated EoS ($P=P(\epsilon)$) using $c_e^2 = dP/d\epsilon$.

The adiabatic speed of sound ($c_s$) is given by energy density derivative of pressure, keeping the particle fractions fixed
\begin{equation}
    c_s^2 = \left(\frac{\partial P}{\partial \epsilon}\right)_{x_i}~.
\end{equation}
Here, $x_i$ represents the particle fractions of various species under consideration. Using Gibbs-Duhem relation (Eq.~\ref{eqn:pressure}) this reduces to
\begin{equation}
    c_s^2 = \frac{1}{\mu_b}\left(\frac{\partial P}{\partial n_b}\right)_{x_i}~.
\end{equation}
Here, $\mu_b$ and $n_b$ are the baryon chemical potential and number density, respectively. The particle fractions are held constant while taking the derivative, and the $\beta$-equilibrium condition is applied after the differentiation. To evaluate this, we need to define the pressure.

The total pressure of DM admixed matter is given by
\begin{equation}
    P = P_{nuc} + P_{l} + P_{\chi}~,
\end{equation}
where $P_{nuc}$ comes from RMF EoS which models hadronic matter and $P_{l}$ is the leptonic contribution from free electrons and muons. The pressure $P_{\chi}$ of the DM component contains two contributions: one from the free fermi gas pressure and the second coming from self-interaction. 

We denote nucleons (neutrons and protons) by $N$, leptons by $l$, and DM by $\chi$. From Gibbs-Duhem relation (Eq.~\ref{eqn:pressure}), we have
\begin{align}
    P &= P_{nuc} + P_{l} + P_{\chi} \\
        &=\sum_N \mu_N n_N + \sum_l \mu_l n_l + \mu_\chi n_\chi - \epsilon_{nuc} -\epsilon_l -\epsilon_\chi~. \\
\end{align}
To calculate $c_s^2$, we need to compute $\partial P/\partial n_b$ keeping the particle fractions fixed. \ Thus,
\begin{align*}
    \left(\frac{\partial P}{\partial n_b}\right)_x &= \sum_N \left(\mu_N\frac{\partial n_N}{\partial n_b} + n_N\frac{\partial \mu_N}{\partial n_b}\right) \\ &+ \sum_l \left(\mu_l\frac{\partial n_l}{\partial n_b} + n_l\frac{\partial \mu_l}{\partial n_b}\right) \\ &+ \mu_\chi\frac{\partial n_\chi}{\partial n_b} + n_\chi\frac{\partial \mu_\chi}{\partial n_b} \nonumber \\
    &- \sum_N \frac{\partial \epsilon_{nuc}}{\partial n_N}\frac{\partial n_N}{\partial n_b} - \sum_l \frac{\partial \epsilon_l}{\partial n_l}\frac{\partial n_l}{\partial n_b} - \frac{\partial \epsilon_\chi}{\partial n_\chi}\frac{\partial n_\chi}{\partial n_b} \\
     &= \sum_N \left(\mu_Nx_N + n_N\frac{\partial \mu_N}{\partial n_b}\right) \\ &+ \sum_l \left(\mu_lx_l + n_l\sum_j x_j\frac{\partial \mu_l}{\partial n_j}\right) \\ &+ \mu_\chi x_\chi + n_\chi x_\chi \frac{\partial \mu_\chi}{\partial n_\chi} \nonumber \nonumber \\
    &- \sum_N \mu_Nx_N - \sum_l \mu_lx_l - \mu_\chi x_\chi \\ 
    &= \sum_N n_N\frac{\partial \mu_N}{\partial n_b} + \sum_l n_lx_l\frac{\partial \mu_l}{\partial n_l} + n_\chi x_\chi \frac{\partial \mu_\chi}{\partial n_\chi}~,
\end{align*}
which can be written as 
\begin{equation}
    \left(\frac{\partial P}{\partial n_b}\right)_x = \left(\frac{\partial P_{nuc}}{\partial n_b}\right)_x + \left(\frac{\partial P_{l}}{\partial n_b}\right)_x + \left(\frac{\partial P_{\chi}}{\partial n_b}\right)_x~.
\end{equation}
As there are no cross-terms between nucleons and leptons or DM, the separation of these terms is justified. The chemical potentials of nucleons only depend on nucleon densities. Chemical potential of each lepton depends on the densities of that particular lepton itself, i.e., $\partial \mu_l / \partial n_j = \delta_{lj} \partial \mu_l / \partial n_l$. Hence, the summation over $j$ drops out. Here, we have defined particle fraction for each species $i$ as $x_i = n_i/n_b$ and used the relation $\partial \epsilon/\partial n_i = \mu_i$. We calculate each of these terms below.

\subsubsection{Nucleonic component}
In the RMF model, the nucleon chemical potentials depend on the densities of all nucleons via the mean value of mesonic fields $\sigma, \omega, \rho$. The nucleon chemical potential is given by
\begin{equation}
    \mu_i = E_{F_i}^* + g_\omega \bar{\omega} + \frac{g_\rho}{2}I_{3i}\bar{\rho}~.
\end{equation}
Here, $I_{3i}$ are components of Pauli matrices and is $+(-)1$ for neutrons (protons). We need to compute $\frac{\partial \mu_i}{\mu_b}$ for nucleons. From Eq. 48 of~\cite{Tran2023}, we have
\begin{align}\label{eqn:delmu_delnb}
    \frac{\partial\mu_i}{\partial n_b}\bigg|_x = 
    \frac{\partial E_{F_i}^*}{\partial n_b}\bigg|_x + \frac{\partial \mu_i^{(m)}}{\partial n_b}\bigg|_x~.
\end{align}

The first term can be solved as follows (Eqs.~50, 51 of~\cite{Tran2023})
\begin{align}
    \frac{\partial E_{F_i}^*}{\partial n_b}\bigg|_x &= 
    \frac{\pi^2x_i}{k_{F_i} E_{F_i}^*} + \frac{ m^*}{E_{F_i}^*}\frac{\partial m^*}{\partial n_b}\bigg|_x~,\\
    \\
    \frac{\partial m^*}{\partial n_b}\bigg|_x  &= - g_{\sigma} 
    \frac{\partial \sigma}{\partial n_b}\bigg|_x~.
\end{align}
We define $m' := \frac{\partial m^*}{\partial n_B}\big|_x$. To evaluate this, we need to solve the Eq.~54. of~\cite{Tran2023}, which is derived from the equation of motion of the scalar field.
\begin{equation} 
\begin{aligned}
    0 &=  
    \frac{m'}{g_\sigma}
    \left(m_\sigma^2 + \frac{\partial^2 U}{\partial \sigma^2}\right) + 
    \sum_i \frac{g_{\sigma}n^s_im' }{m^*}  \\
    &+ \sum_i 
    g_{\sigma}
    \frac{m^*}{2\pi^2}
    \bigg[
    \frac{\pi^2 x_i}{k_{F_i}^2}  E_{F_i}^* 
    + k_{F_i}\left(\frac{\pi^2x_i}{k_{F_i} E_{F_i}^*} + \frac{ m^*}{E_{F_i}^*}m'\right)\bigg]\\
    &\quad  + \sum_i g_{\sigma} \frac{m_i^*}{2\pi^2}
    \bigg[
    -2m^*m'
    \ln\frac{k_{F_i} + E_{F_i}^*}{m^*}\\
    &\qquad \quad \quad 
    - {m^*}^2
    \left[\frac{\dfrac{\pi^2 x_i}{k_{F_i}^2} + \dfrac{\pi^2x_i}{k_{F_i} E_{F_i}^*} + \dfrac{ m^*}{E_{F_i}^*}m' }{k_{F_i} + E_{F_i}^*} -
        \frac{m'}{m_i^*}\right]\bigg] ~.
        \label{eqn:d_sigma_eom}
\end{aligned}
\end{equation} 

Solving for $m'$ we get,
\begin{align}
    m' &= \frac{\partial m^*}{\partial n_B}\big|_x \\
    &= -\left(\frac{g_\sigma}{m_\sigma}\right)^2m^*\frac{\sum_i\dfrac{x_i}{E_{F_i}^*}}{\left(1 + \dfrac{1}{m_\sigma^2}\dfrac{\partial^2 U}{\partial \sigma^2} + \dfrac{1}{\pi^2}\left(\dfrac{g_\sigma}{m_\sigma}\right)^2r \right)}
\end{align}
where $r$ is
\begin{equation}
    r = \sum_i \frac{1}{2}k_{F_i} E_{F_i}^* - \frac{3}{2}m^{*2}\ln\frac{k_{F_i} + E_{F_i}^*}{m^*} + m^{*2}\frac{k_{F_i}}{E_{F_i}^*}~.
\end{equation}

The second term of Eq.~\ref{eqn:delmu_delnb} can be evaluated by solving the system of linear equations as described in Sec. V. B of~\cite{Tran2023}. This term $\mu_i^{(m)}$ consists of the terms $\mu_i^{(m)} =  g_\omega \bar{\omega} + \frac{g_\rho}{2}I_{3i}\bar{\rho}$. Thus,
\begin{equation}\label{eqn:mu_i^m}
    \frac{\partial \mu_i^{(m)}}{\partial n_B}\bigg|_x = g_{\omega}\frac{\partial\omega}{\partial n_B}\bigg|_x + \frac{1}{2}I_{3i}g_{\rho}\frac{\partial \rho}{\partial n_B}\bigg|_x~.
\end{equation}
The equations of motion for these mesonic fields $\bar{\omega}$ and $\bar{\rho}$ are given in Eq.~\ref{mesonic_eoms}.
We now take the derivative with respect to $n_b$ and denote this by prime. Thus,
\begin{align}
    (m_\omega^2 + 2 \Lambda_\omega g_{\rho N}^2 g_{\omega N}^2 \rho^2) \omega' + (4 \Lambda_\omega g_{\rho N}^2 g_{\omega N}^2 \omega \rho) \rho' &=   g_{\omega} (x_p + x_n)~, \\
    (m_\rho^2  + 2 \Lambda_\omega g_{\rho N}^2 g_{\omega N}^2 \omega^2) \rho' +  (4 \Lambda_\omega g_{\rho N}^2 g_{\omega N}^2 \omega \rho) \omega' &=  \frac{g_{\rho}}{2} (x_p-x_n)~. 
\end{align}
This system of equations linear in $\omega'$ and $\rho'$ can be solved simultaneously to get the solutions. These can then be used in Eqn.~\ref{eqn:mu_i^m} to get $\partial \mu_i^{(m)} /\partial n_B)_x$ which can be further used in Eqn.~\ref{eqn:delmu_delnb} to get the second term of $(\partial\mu_i/\partial n_b)_x$. Finally, $\sum_N n_N (\partial\mu_N /\partial n_b)_x$ gives the nucleonic contribution to $(\partial P/\partial n_b)_x$ from which the adiabatic speed of sound is calculated by dividing by $\mu_b$.

\subsubsection{Leptonic component}
We include leptons in the bulk matter which are responsible for charge neutrality and maintaining the $\beta$-equilibrium. 

\textit{Electrons:} \\
The electrons are highly degenerate and ultra-relativistic as their fermi momentum is much larger than electron mass ($\mu_e \approx k_{F_e} \gg m_e$).
Here $\mu_e$ is the electron chemical potential, $k_{F_e}\approx \mu_e$ is the electron Fermi momentum, and the electron number density is given by $n_e = k_{F_e}^3/3\pi^2$. 

\begin{align}
    \left(\frac{\partial P}{\partial n_b}\right)_{e} &= n_ex_e\frac{\partial \mu_e}{\partial n_e} \\
    &= n_ex_e\frac{\partial \mu_e}{\partial k_{F_e}}\frac{\partial k_{F_e}}{\partial n_e} \\
    &= n_ex_e\times 1 \times \frac{\pi^2}{k_{F_e}^2} \\
    &= \frac{\mu_ex_e}{3}~.
\end{align}
Here, we have used $\mu_e \approx k_{F_e}$ and $n_e = k_{F_e}^3/3\pi^2$.

\textit{Muons:}\\
The chemical potential of muons is given by $\mu_{\mu} = \sqrt{k^2_{F_{\mu}} + m_{\mu}^2}$, where $n_{\mu} = k^3_{F_{\mu}}/3\pi^2$.
\begin{align}
    \left(\frac{\partial P}{\partial n_b}\right)_{\mu} &= n_{\mu}x_{\mu}\frac{\partial \mu_{\mu}}{\partial n_{\mu}} \\
    &= n_{\mu}x_{\mu}\frac{\partial \mu_{\mu}}{\partial k_{F_{\mu}}}\frac{\partial k_{F_{\mu}}}{\partial n_{\mu}} \\
    &= n_{\mu}x_{\mu}\times \frac{k_{F_{\mu}}}{\mu_{\mu}} \times \frac{\pi^2}{k_{F_{\mu}}^2} \\
    &= \frac{k_{F_{\mu}}^2x_{\mu}}{3\mu_{\mu}}~.
\end{align}

\subsubsection{DM component}

The chemical potential of DM is given by $\mu_{\chi} = \sqrt{k^2_{F_{\chi}} + m_{\chi}^2} + Gn_{\chi}$, where $n_{\chi} = k^3_{F_{\chi}}/3\pi^2$.
We can calculate the DM component as follows:
\begin{align}
    \left(\frac{\partial P}{\partial n_b}\right)_{\chi} &= n_\chi x_\chi\frac{\partial \mu_\chi}{\partial n_\chi} \\
    &= n_\chi x_\chi \frac{\partial \sqrt{k_{F_\chi}^2+m_\chi^2}}{\partial k_{F_\chi}}\frac{\partial k_{F_e}}{\partial n_e} +  n_\chi x_\chi G\\
    &= n_\chi x_\chi \times \frac{k_{F_\chi}}{E_{F_\chi}} \times \frac{\pi^2}{k_{F_\chi}^2} +  n_\chi x_\chi G\\
    &=\frac{k_{F_{\chi}}^2x_{\chi}}{3E_{F_\chi}} + n_\chi x_\chi G~.
\end{align}
Here we have used $\mu_\chi = E_{F_\chi} + G n_\chi = \sqrt{k_{F_\chi}^2+m_\chi^2} + G n_\chi$ and $n_\chi = k_{F_\chi}^3/3\pi^2$.

\section{Results}
\label{sec:results}



\begin{figure*}
    \centering
    \includegraphics[width = 0.5\linewidth]{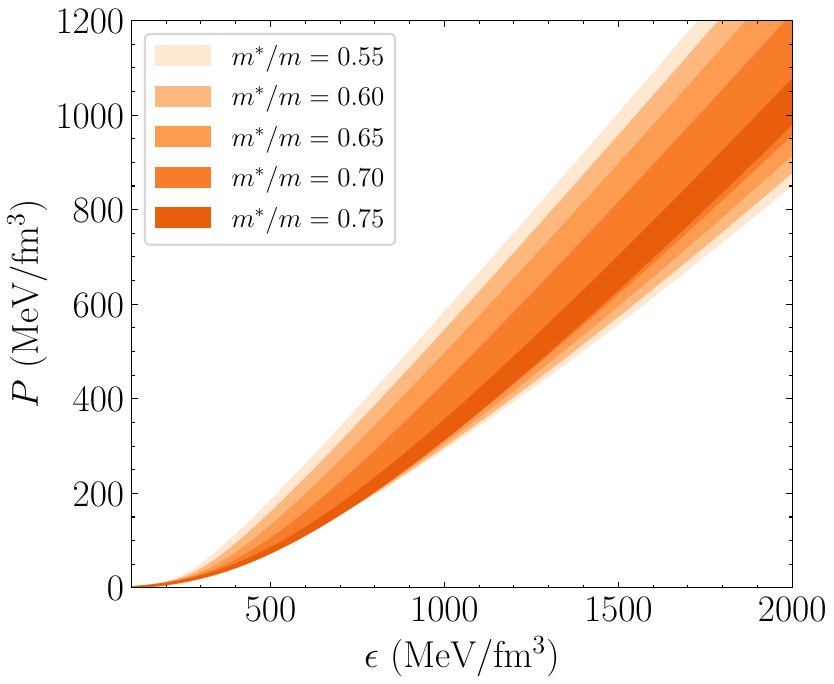}
    \includegraphics[width = 0.49\linewidth]{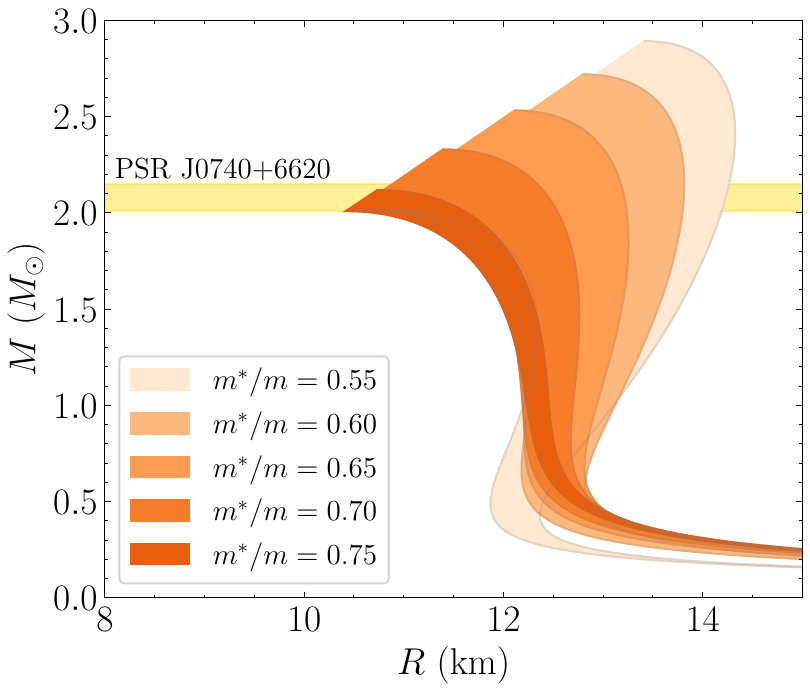}
    \caption{a) The EoS bands and corresponding b) mass-radius bands for DM admixed NS for ``RMF" set parameters satisfying the two solar mass constraint. The widest (faintest) band corresponds to $m^*/m=0.55$, and the narrowest (darkest) corresponds to $m^*/m=0.75$. The intermediate ones are for increasing values of $m^*/m$ in steps of 0.05 as we go from the widest to the narrowest. See text for more details. 
    }
    \label{fig:eos_mr}
\end{figure*}

 In Fig.~\ref{fig:eos_mr}a, we display the bands of EoSs ($P$ vs $\epsilon$) for nuclear matter admixed with DM for the different nucleon effective masses considered in the ``RMF" set. The effect of effective mass can be seen by considering only the top boundary curve of each EoS band. 
 For each $m^*/m$, the width of each orange band is obtained by varying $G$, satisfying the two solar mass constraint, which gives the effect of dark matter on the EoS. The $M-R$ bands corresponding to these EoS bands are shown in Fig.~\ref{fig:eos_mr}b. Here, again, the stiffer boundary for each band gives the $M-R$ curve without DM, marking the effect of $m^*/m$, and the thickness of the corresponding bands shows the effect of DM in the $M-R$ space. The faintest and the broadest band corresponds to $m^*/m=0.55$. The upper edge of the band (stiffest) corresponds to pure nucleonic EoSs ($f_{DM} = 0$), and the lower edge (softest) to the one with the maximum DM fraction that satisfies the two solar mass constraint. The inclusion of DM is known to soften the EoS and lower the maximum TOV mass~\citep{motta2018a, motta2018b, Shirke2023b}. The next darker band is the same, but for a softer nuclear EoS with $m^*/m=0.60$. We see the band is narrower and gets truncated on both the stiff and soft ends. The next consecutive dark bands correspond to $m^*/m=0.65$, $m^*/m=0.70$, and $m^*/m=0.75$ (most narrow and dark) respectively. We will study the $g$-modes corresponding to these bands in the next section. As a special case, we also consider the ``Stiffest" EoS (see Table.~\ref{table:parameters} parametrization for a few cases. This is the stiffest RMF EoS satisfying the low-density CEFT constraints~\citep{Ghosh2022EPJA}. The band corresponding to this case is the broadest and almost similar to that of $m^*/m=0.55$ and is not shown in the figure.

\subsection{Effect of DM}

\begin{figure}
    \centering
    \includegraphics[width = \linewidth]{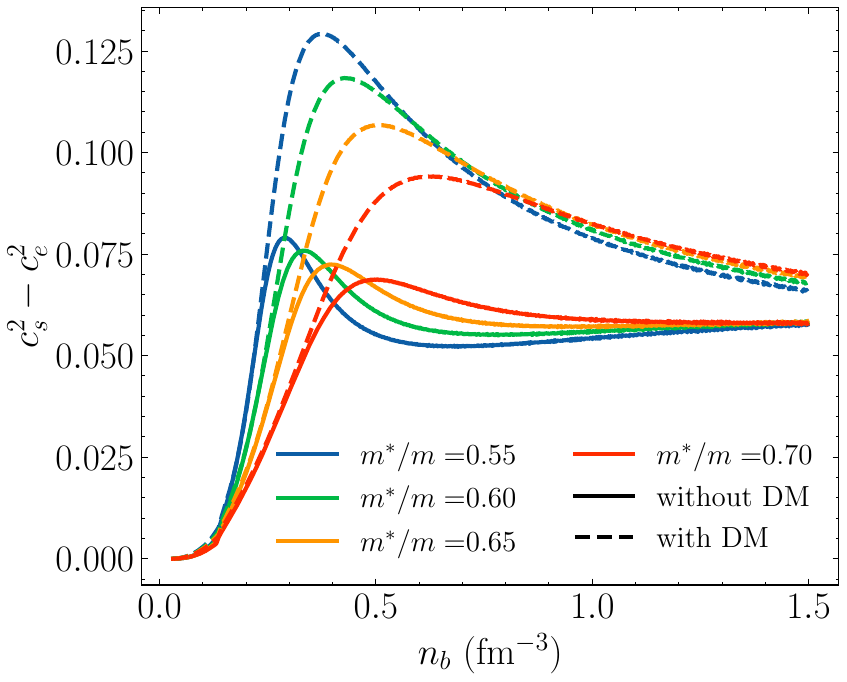}
    \caption{Sound speed difference $c^2_s-c^2_e$ plotted for the various effective mass values of the ``RMF" set. The solid curves are for the EoSs without any DM and the dashed ones are for DM admixed EoSs with $G=30$ fm$^2$.} 
    \label{fig:speed_difference}
\end{figure}

In Fig.~\ref{fig:speed_difference}, we plot the difference in the adiabatic and equilibrium speeds of sound for the ``RMF" set with and without DM. The solid curves correspond to purely nucleonic EoSs, whereas the dashed ones include DM, where the DM self-interaction parameter $G$ is kept fixed to $30$ fm$^2$. For $G=30$ fm$^2$, the EoS corresponding to $m^*/m=0.75$ is soft and does not result in a two-solar mass NS configuration, so we do not include the $m^*/m=0.75$ case. We see the speed difference is consistently higher in the case of DM admixed matter, wherein we have an additional particle species. A rise in the speed difference has also been reported in the case of hyperonic matter beyond the density threshold at which hyperons appear~\citep{Tran2023}. In the case of DM admixed NSs, DM is present at all densities within the NS core, and the DM fraction increases with density. Here, we observe that the difference between the case with and without admixture of DM grows with density, and only starts to reduce at very high densities. In summary, the curves for the DM admixed case differ from those without any DM, as follows:
\begin{itemize}
    \item The speed difference is larger at all densities. This can be attributed to the additional degree of freedom, the DM particle, as also observed in the case of the appearance of hyperon and quark degrees of freedom~\citep{Tran2023, Jaikumar:2021jbw}. Consequently, the peak value of the curves with DM is higher.
    \item The curves peak at higher densities. This could be a result of softening due to the inclusion of DM. As can be seen in the case without DM, the curves for higher effective mass, i.e., softer EoSs, peak at higher densities.
    \item The curves with DM do not have a minima in the considered density range. This, again, could be attributed to the softening of EoS. As can be seen in the case without DM, the curves have a sharper dip after the maxima for stiff EoSs and become flatter as we go to softer EoSs. All the EoSs with DM here are relatively softer.
\end{itemize}

We now investigate the effect of the self-interaction parameters $G$ on the resultant \gms, which is the only DM microscopic parameter of this model.

\subsection{Effect of DM self-interaction $G$}

The $g$-mode frequency can be computed using the adiabatic and equilibrium sound speeds, as outlined in the previous section. We first compute it for the case of fixed nuclear EoS. We consider ``RMF" parameterization with $m^*/m=0.65$. 

\begin{figure*}
    \centering
    \includegraphics[width = 0.49\linewidth]{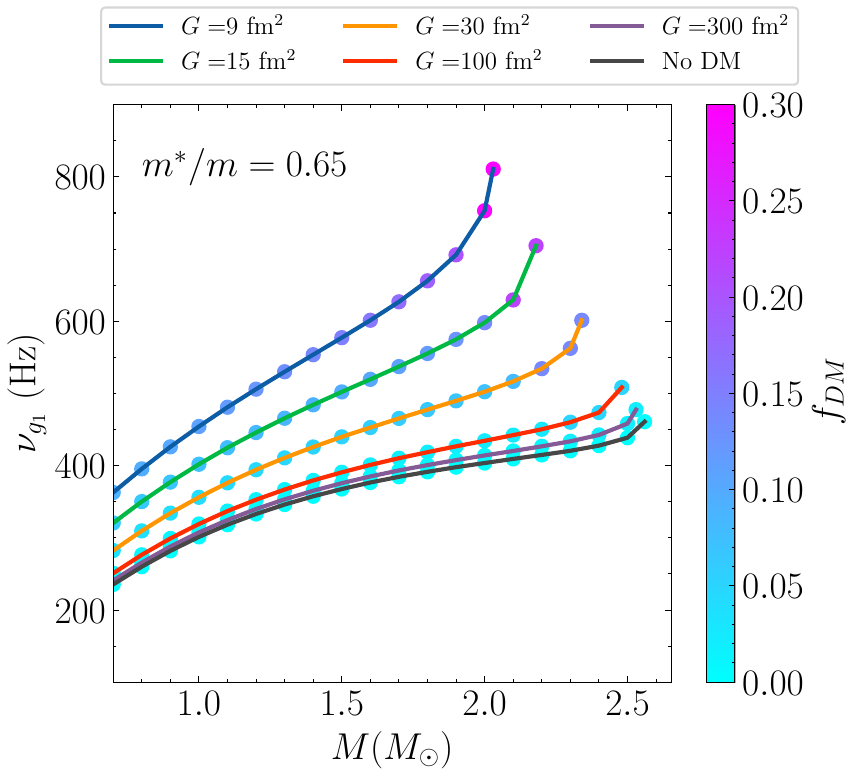}\label{fig:gmodes_m0.65_g1}
    \includegraphics[width = 0.49\linewidth]{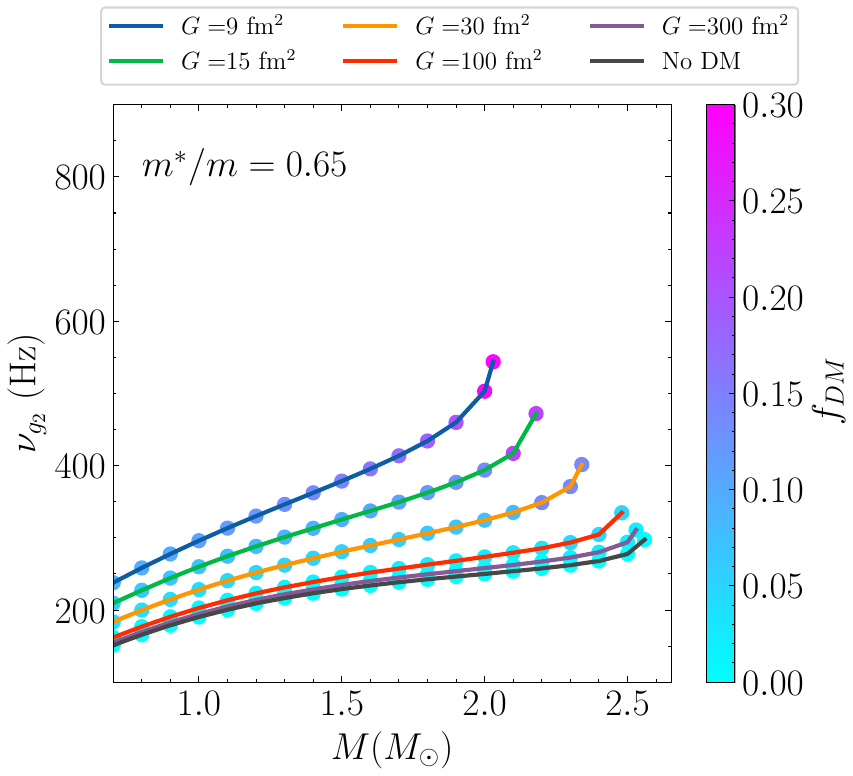}\label{fig:gmodes_m0.65_g2}
    \caption{The fundamental mode a) $g_1$ and first overtone b) $g_2$ of $g$-mode frequency as a function of mass for DM admixed NS with hadronic matter modeled by ``RMF" set of parameters as enlisted in~\ref{table:parameters}. $m^*/m$ has been fixed to 0.65. We show the variation of $\nu_{g_1}$ and $\nu_{g_2}$ for different values of $G$. The colour bar denotes the DM fraction.} 
    \label{fig:gmodes_m0.65}
\end{figure*}

In Fig.~\ref{fig:gmodes_m0.65} a), we show the principal mode ($g_1$)  of the $g$-mode frequency for purely hadronic EoS (black), i.e., $f_{DM}=0$ (denoted by  'No DM') and for DM admixed NSs for varying values of $G$. In Fig.~\ref{fig:gmodes_m0.65} b), we show the corresponding first overtone $g_2$. We show the DM fraction for the DM admixed NS for different mass configurations using a colour bar. Note that the relation of $f_{DM}$ with NS mass also depends on $G$ and weakly on the underlying nuclear EoS. This was shown in one of our earlier works~\citep{Shirke2024}. Hence, we do not show $f_{DM}$ on the opposite axis as that of mass. The DM fraction for the black curve (purely hadronic) is zero throughout.

We make the following observations:
\begin{itemize}
    \item The frequency for both $g_1$ and $g_2$ increases with mass in all the cases. This is expected as we saw in Fig.~\ref{fig:speed_difference} that the speed difference and hence the \bv~frequency (Eq.\ref{eqn:brunt_vaisala_frequency}), that dictates the $g$-mode frequency scale, increases on the inclusion of DM.
    \item $g_1 > g_2$ in both cases, as expected (discussed in Sec I: Introduction)
    \item There is a systematic increase in the $g$-mode upon inclusion of DM for both harmonics throughout the mass range. This can be understood from the fact that DM is present throughout the core EoS, even at low densities. There is no critical density beyond which DM appears, as in the case of quarks or hyperons.
    \item The $g$-mode frequency increases with decreasing $G$ (or increasing DM fraction). This is because the speed difference is higher when the DM fraction is higher.
\end{itemize}
Since the $g$-mode frequency increases with mass and decreases with $G$, we expect it to increase with $f_{DM}$. This is because, as illustrated in~\cite{Shirke2024}, $f_{DM}$ is a monotonically increasing function of $(M/G)$. 

\begin{figure}
    \centering
    \includegraphics[width = \linewidth]{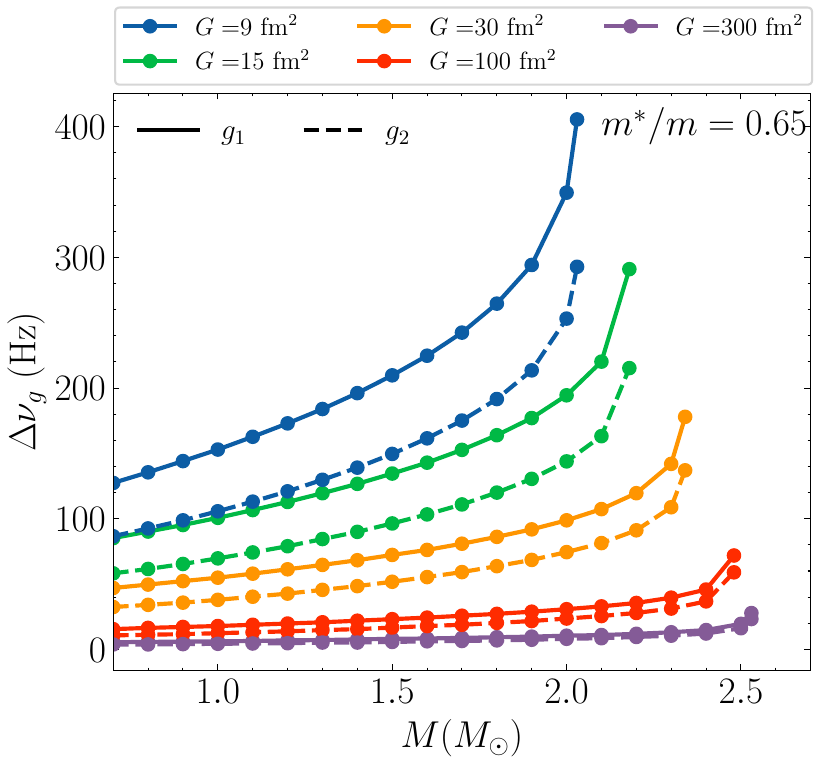}
    \caption{The increase in fundamental $g1$ and first overtone $g_2$ of $g$-mode frequency as a function of mass for DM admixed NS with hadronic matter modelled by ``RMF" set of parameters as enlisted in~\ref{table:parameters}. $m^*/m$ has been fixed to 0.65. We show the variation of $\nu_{g_1}$ (solid) and $\nu_{g_2}$ (dashed) for different values of $G$. } 
    \label{fig:gmodes_delta_g_m0.65}
\end{figure}

The change in the $g$-mode frequency with and without DM is defined by $\Delta \nu_g (G, M) \equiv \nu_g(G, M) - \nu_g(G \to \infty, M)$. Here, the dependence on the nuclear parameters, such as the effective mass considered here, is implicit. $\Delta \nu_g (G, M)$, in general, would also depend on the effective mass considered. In Fig.~\ref{fig:gmodes_delta_g_m0.65}, we plot the increase in the fundamental $g$-mode frequency $g_1$ (a) and first overtone $g_2$ (b)  corresponding to the configurations considered in Fig.~\ref{fig:gmodes_m0.65}. We observe the following
\begin{itemize}
    \item $\Delta \nu_g$ is positive throughout. So, the inclusion of DM raises the $g$-mode frequency
    \item $\Delta \nu_g$ is higher for lower $G$ (higher DM fraction)
    \item $\Delta \nu_g$ is higher for $g_1$ as compared to $g_2$ for all the masses.
    \item $\Delta \nu_g$ increases with NS mass, so in addition to a systematic rise in \gms~in Fig.~\ref{fig:gmodes_m0.65} throughout the mass range, this rise has a slight dependence on NS mass.
\end{itemize}
 We observe that $\Delta \nu_g$ also increases with $M$ and falls with $G$. Hence, we expect it to be a rising function of $f_{DM}$ as well. Alternatively, we can also write $\Delta \nu_g (G, M) \equiv \nu_g(f_{DM}, M) - \nu_g(f_{DM} = 0, M)$. This can be justified, as an empirical universal relation was found in an earlier work~\citep{Shirke2024}, where it was shown that the DM fraction only depends on $G$ and the mass of the DM admixed NS configuration and is independent of nuclear saturation parameters. We explore the dependence of $\Delta \nu_g$ on $f_{DM}$ in Sec~\ref{sec:deltag_dmfrac}.

It would be useful to incorporate the effect of the underlying nuclear EoS on $g$-modes to delineate the effect of DM. The effect of nuclear saturation parameters on NS macroscopic properties has been widely studied for RMF EoSs and it has been shown that the effective nucleon mass is the most dominant parameter that dictates the NS observables~\citep{Pradhan2021, Ghosh2022EPJA,Ghosh2022FrASS, Shirke2023a, Maiti2024}. The effective nucleon mass is responsible for controlling the EoS stiffness at high densities. Hence, we only vary the effective nucleon mass to study the uncertainty in the nuclear EoS. The usual range of the Dirac effective mass for RMF EoSs is $0.55 \le m^*/m \le 0.75$, where the lower values correspond to a stiff EoS and higher for soft. The corresponding bands for DM admixed EoS were shown in Fig.~\ref{fig:eos_mr}a for select values of $m^*/m$. 

\subsection{Effect of effective nucleon mass $m^*/m$}

\begin{figure*}
    \centering
    \includegraphics[width = 0.49\linewidth]{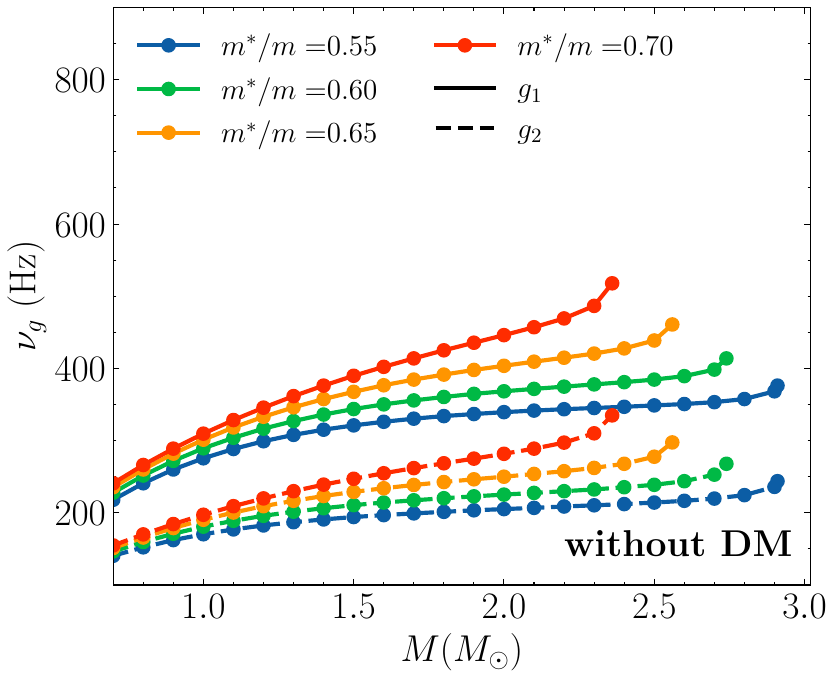}\label{fig:gmodes_gm1_withoutdm}
    \includegraphics[width = 0.49\linewidth]{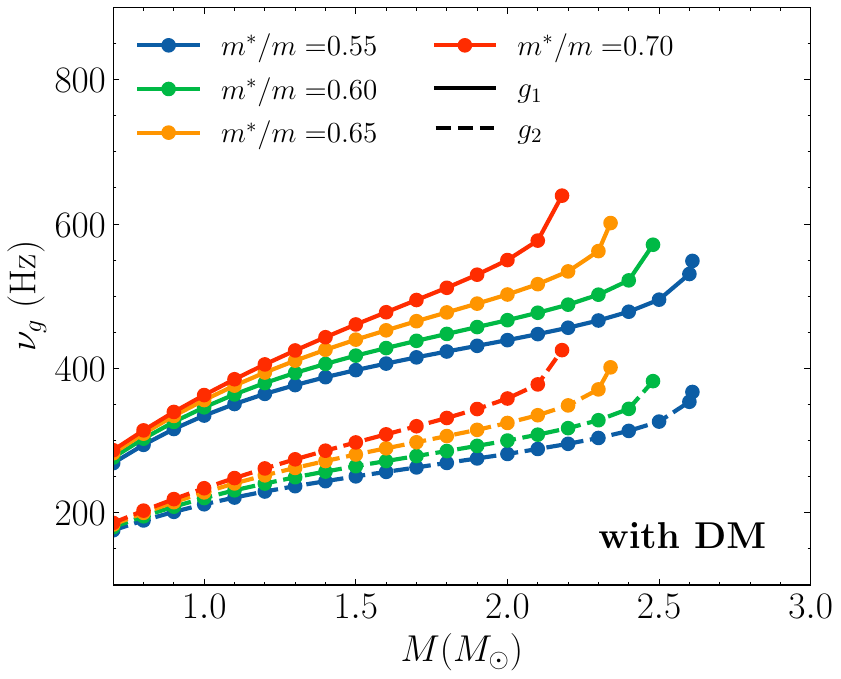  }\label{fig:gmodes_gm1_withdm}
    \caption{The fundamental $g1$ and first overtone $g_2$ of $g$-mode frequency for a) pure hadronic NS with ``RMF" EoS and b) DM admixed NS with hadronic matter modeled by the RMF EoS and the DM self-interaction strength set to $G=30$ fm$^2$. This has been shown for the different effective masses of the RMF EoS in the range [0.55,0.75]. As $G=30$ fm$^2$, the DM fraction is approximately given by $f_{DM} = \frac{1}{30}(\frac{M}{M_{\odot}})$ using the relation proposed in~\citet{Shirke2024}. } 
    \label{fig:gmodes_vary_mstar}
\end{figure*}

To study the effect of the in-medium nucleon mass on $g$-modes, we fix the DM parameter $G$ to $30$ fm$^2$. The corresponding EoS for $m^*/m=0.75$ is too soft to satisfy the two solar mass constraint; hence, we do not show it here. Fig.~\ref{fig:gmodes_vary_mstar} shows the $g_1$ and $g_2$ frequencies (a) without DM and (b) with DM for fixed $G=30$ fm$^2$, respectively, for select values of effective mass $m^*/m=0.55,0.60,0.65,0.70$. These configurations correspond to those whose sound speed difference was shown in Fig.~\ref{fig:speed_difference}. As seen before, the frequencies upon inclusion of DM are higher. Qualitatively, we see the effect of effective mass to be similar in both the presence and absence of DM. We observe that the $g$-mode frequency is higher for lower effective nucleon mass. This is as expected, as we know that higher effective mass softens the EoS, resulting in a higher $g$-mode frequency. The range for $g_1$ and $g_2$ with and without DM corresponds to the range of considered effective nucleon mass tabulated in Table.~\ref{table:g_range_vary_mstar}. The ranges for both $1.4 M_{\odot}$ and $2 M_{\odot}$ DM admixed NS configurations are provided.

\begin{table}
    \centering
    \begin{tabular}{c|c|c|c}
         Mass &  Composition & $\nu_{g_1}$ (Hz) & $\nu_{g_2}$ (Hz)\\
         \hline
         \hline
         1.4 $M_{\odot}$ & with DM & $387.9 - 443.2$ &$ 243.8 - 286.0$\\
         & without DM & $315.1 - 376.2$ & $190.85 - 239.0$\\
         \hline
         2.0 $M_{\odot}$ & with DM & $439.2 - 550.1$ & $281.5 - 358.3$\\
         & without DM & $339.4 - 446.1$ & $205.2 - 281.8$\\
    \end{tabular}
    \caption{Range of $g$-mode frequencies for $1.4 M_{\odot}$ and $2 M_{\odot}$ DM admixed NSs with fixed $G=30$ fm$^{2}$ and effective mass in the range [0.55, 0.70]. }
    \label{table:g_range_vary_mstar}
\end{table}

\begin{figure}
    \centering
    \includegraphics[width = \linewidth]{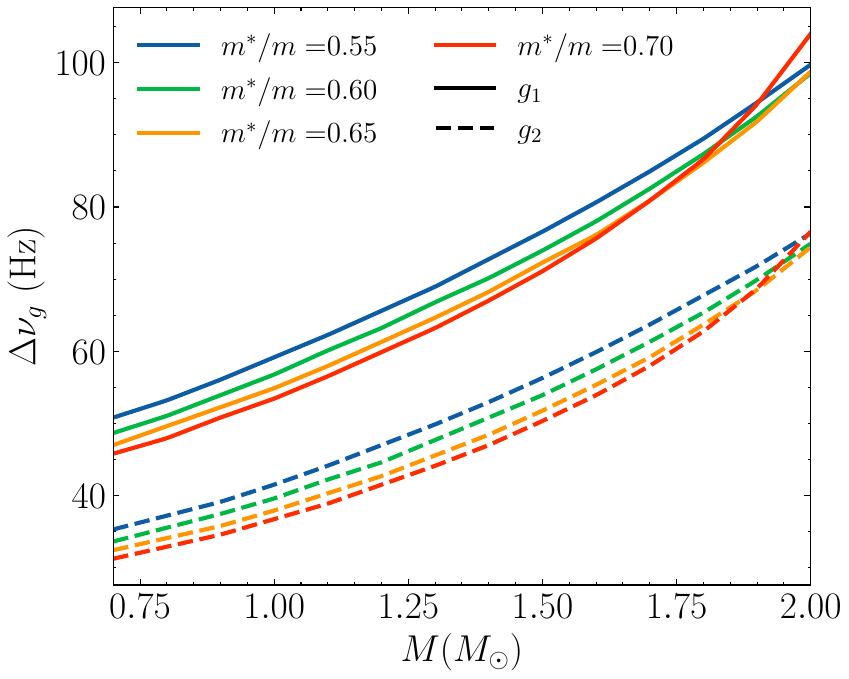}
    \caption{The increase in fundamental $g_1$ and first overtone $g_2$ of $g$-mode frequency for a DM-admixed NS with ``RMF" EoS when DM with self-interaction strength set to $G=30$ fm$^2$ is added. This is shown for different values of effective mass $m^*/m$ for RMF EoS in the range [0.55,0.75] } 
    \label{fig:delta_gmodes_vary_mstar}
\end{figure}

For all the cases considered in Fig.~\ref{fig:gmodes_vary_mstar}, the rise in the $g$-mode frequency as compared to the case without DM is plotted in Fig.~\ref{fig:delta_gmodes_vary_mstar}. Again, we see that $\Delta \nu_g$ is higher in the case of $g_1$ and that it increases with mass. The difference here this time is that $G$ is fixed. From the figure, we can see that below $2 M_{\odot}$, $\Delta \nu_g$ is higher for lower effective masses. Thus, the inclusion of DM for a fixed $G$ has a greater effect on the stiffer EoSs than on the soft ones. However, we note that the effect of stiffness is not prominent. The range of rise in frequencies for different effective masses is within a range of a few 10 Hz as compared to the actual $g$-mode frequencies, which are  $\mathcal{O}(100)$ Hz. We saw in Fig.~\ref{fig:gmodes_delta_g_m0.65} that $\Delta \nu_g$ depends strongly on the value of $G$ and spanned a range of a few 100 Hz.  Thus, the range of $\Delta \nu_g$ coming from the nuclear uncertainty for a given NS mass configuration is much smaller as compared to that coming from the uncertainty in DM interactions. Thus, the effect of DM on \gms~is very prominent, and cannot be ignored. We conclude that the rise in $g$-mode frequency is much more strongly correlated with uncertainties in the dark sector than those in the nuclear sector. We now explore this dependence in detail.

\subsection{DM fraction and $g$-modes}\label{sec:deltag_dmfrac}

It is more natural to think in terms of the DM fraction for a given configuration of DM admixed NS instead of the microscopic DM parameter $G$. The relation of $f_{DM}$ with NS mass and $G$ was provided in~\cite{Shirke2024}. Therein, it was shown that it is a monotonic function of $(M/G)$. Here, for a given configuration of DM admixed NS, we calculate the DM fraction as given in Eq.~\ref{eqn:dm_frac}. 

\begin{figure}
    \centering
    \includegraphics[width = \linewidth]{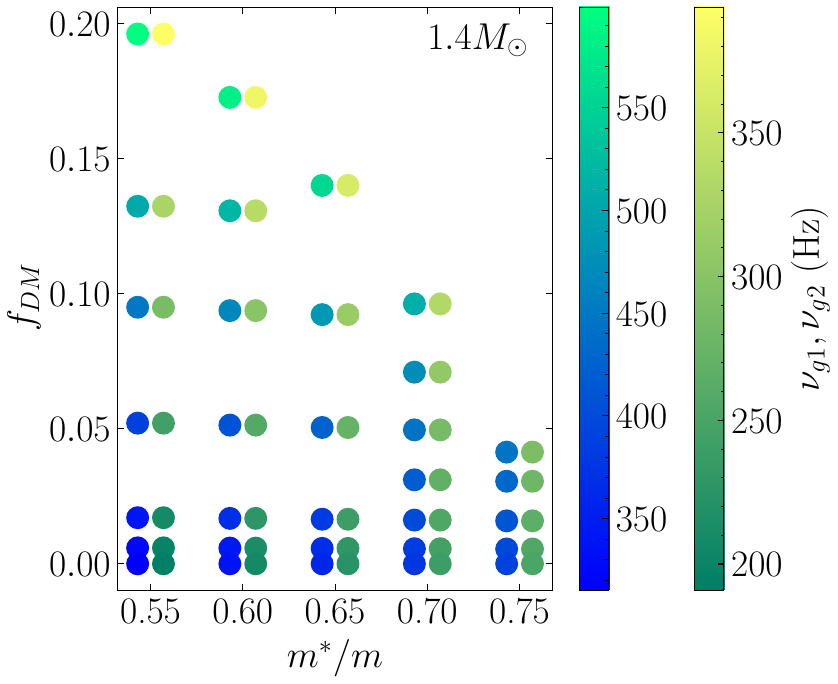}
    \caption{$g_1$ (blue colour bar) and $g_2$ (green colour bar) frequencies for various effective mass values $m^*/m \in \{0.55,0.60,0.65,0.70,0.75\}$ and select values of DM fraction. The mass of the star is fixed to $1.4 M_{\odot}$.} 
    \label{fig:freq_fdm_mstar_1.4}
\end{figure}

In Fig.~\ref{fig:freq_fdm_mstar_1.4}, we show the $g_1$ and $g_2$ frequencies for varying effective nucleon mass and DM fraction for cases considered here. In this plot, the $g_1$ and $g_2$ frequencies are shown as pairs using two separate colour bars. In each pair, the left spot corresponds to $\nu_{g1}$ and the right to $\nu_{g2}$. The first and the second colour bars correspond to $\nu_{g1}$ and $\nu_{g2}$, respectively. The values of nucleon effective mass considered are 0.55, 0.60, 0.65, 0.70, 0.75. For each of these values, we have plotted a pair for select values of the DM fraction. Each pair corresponds to a fixed value of $G$. This plot only shows the case for $1.4 M_{\odot}$ DM admixed NS. We can see that $f_{DM}$ in this case goes up to $\sim0.2$. For a fixed DM admixed NS mass, we see that both $\nu_{g_1}$ and $\nu_{g_2}$ strongly rise with $f_{DM}$. We also observe a mild dependence on the effective nucleon mass, as the frequencies increase with increasing $m^*/m$. However, the effect of $f_{DM}$ is more prominent. Thus, we can conclude that $\nu_{g_i} = \nu_{g_i}(M, f_{DM}, m^*/m)$, with the dependence on $f_{DM}$ being stronger than that on $m^*/m$. Here, $i=1,2$ correspond to the fundamental and the first overtone of $g$-modes, respectively.

To quantify this dependence, it is important to see the rise in the $g$-mode frequencies introduced by different parameters. As, $\nu_{g_i} = \nu_{g_i}(M, f_{DM}, m^*/m)$, we study $\Delta \nu_{gi}$ for different $M$, $f_{DM}$, and $m^*/m$. Earlier, we had concluded that $\Delta \nu_{g_i}$ is higher for higher NS mass, lower $G$, and lower $m^*/m$. This was done either by keeping $m^*/m$ or $G$ fixed. We also concluded that this implies that $\Delta \nu_{g_i}$ is a rising function of $f_{DM}$, and consider $f_{DM}$ in place of $G$. 

\begin{figure}
    \centering
    \includegraphics[width = \linewidth]{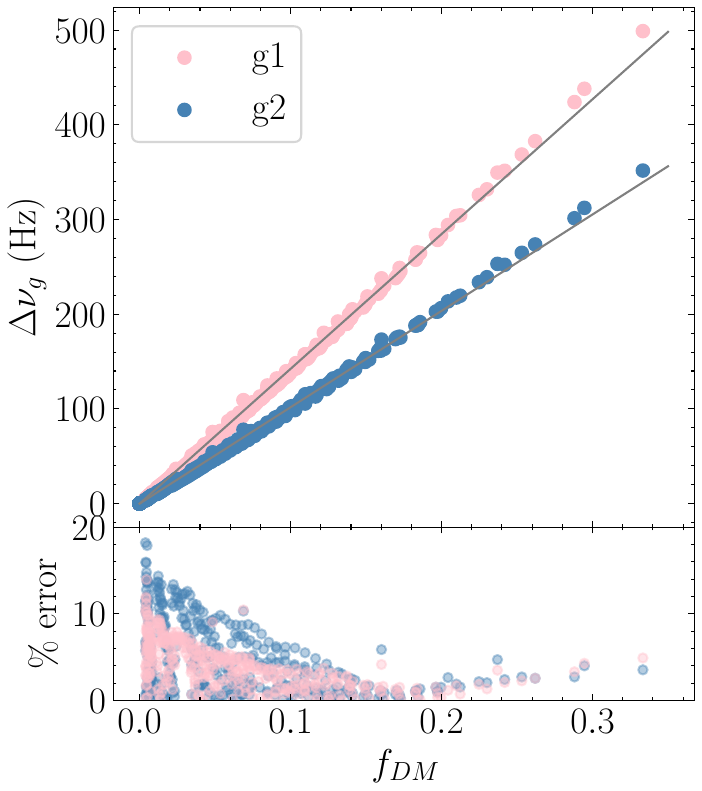}
    \caption{The rise in $g$-mode frequencies $\Delta \nu_g$ for $g_1$ (pink) and $g_2$ (blue), when all the parameters $m^*/m$, $G$ and the total mass of DM admixed NS are varied.} 
    \label{fig:Deltag_fdm}
\end{figure}

Now, we consider the cases where all the parameters, namely, $M$, $f_{DM}$, and $m^*/m$ are varied i.e., we consider all the configurations shown in Fig.~\ref{fig:freq_fdm_mstar_1.4} for $1.4 M_{\odot}$ and also all the NS mass configurations in the range $M \in [0.7, 2] M_{\odot}$ in steps of $0.1 M_{\odot}$. For each of these configurations, we calculated $\Delta \nu_{g_i}(M, f_{DM}, m^*/m) = \nu_{g_i}(M, f_{DM}, m^*/m) - (M, f_{DM}=0, m^*/m)$. We plot $\Delta \nu_{g_i}$ for all these configurations as a function of $f_{DM}$ in Fig.~\ref{fig:Deltag_fdm}. We colour code the scatter plot with pink for $\Delta \nu_{g_1}$ and blue for $\Delta \nu_{g_2}$. It turns out that $\Delta \nu_{g_i}$ only depends on the DM fraction. Note that $f_{DM}$ has implicit mass and $G$ dependence~\citep{Shirke2024}, but we do not find any explicit dependence on these parameters. All the scatter points for each $g_1$ and $g_2$ follow a linear relationship. We fit a straight line for each of these two relations. We report the fit relations here:
\begin{align}\label{eqn:Deltag_fdm_fit}
    \Delta \nu_{g_1} = 14.23(\pm0.02) \times f_{DM}[\%] \\
    \Delta \nu_{g_2} = 10.18(\pm0.02) \times f_{DM}[\%]    
\end{align}

There is a small spread in this fit relation (up to $15\%$ and $20\%$ in $g_1$ and $g_2$, respectively), which is attributed to the weak dependence on the effective mass and the explicit mass dependence. For the fit, we demanded that $\Delta \nu_{g_i} (f_{DM}=0) = 0$, since no rise in $g$-mode is expected if there is no DM. Thus, although the $g$-mode frequencies depend on the DM admixed NS mass, nucleon effective mass and DM fraction, the rise in $g$-mode frequencies follows a quasi-universal relation with DM fraction regardless of the underlying microscopic EoS.

\subsection{Observability}

\begin{figure*}
    \centering
    \includegraphics[width = 0.49\linewidth]{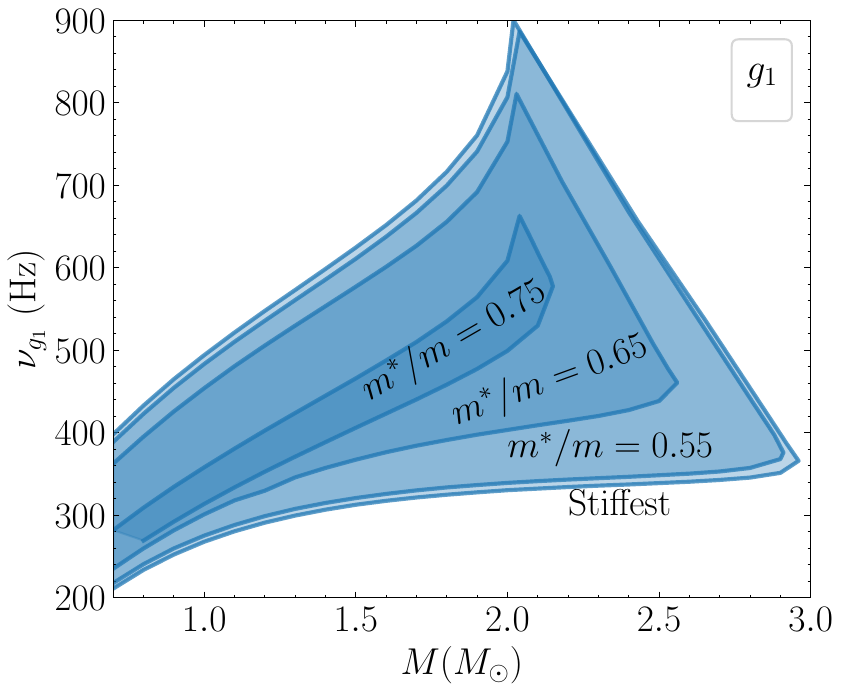}
    \includegraphics[width = 0.49\linewidth]{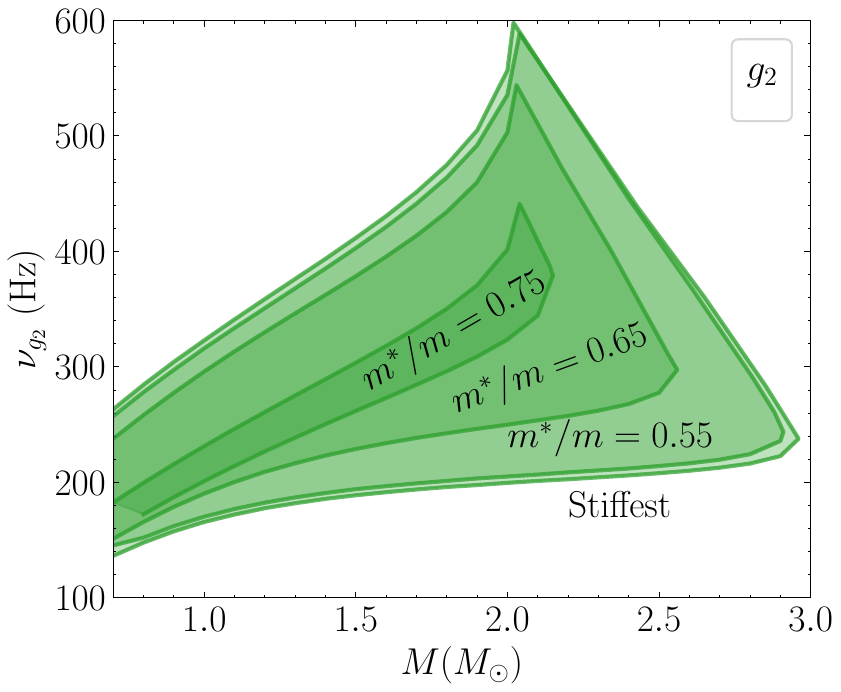}
    \caption{(a) The region spanned by DM admixed NSs considered in this work in the frequency-mass plane for the fundamental $g$-mode frequency $g_1$. The outermost (widest) band is for the stiffest RMF EoS consistent with the chiral effective field theory constraints. The bands for ``RMF" parameterization are also shown for effective nucleon mass values 0.55, 0.65, and 0.75. (b) Same as (a) but for the first overtone $g_2$ of g-mode oscillations.} 
    \label{fig:g1_contours}
\end{figure*}

Finally, we study the area spanned by DM admixed NSs for the DM model considered in this work in the $\nu_g-M$ plane. In Fig.~\ref{fig:g1_contours}, we plot the bands for $g_1$ (a) and $g_2$ (b) frequencies for different nuclear EoSs. In particular, we have shown the bands for $m^*/m=0.55,0.65,0.75$ (corresponding to the EoS and $M-R$ bands shown in Fig.~\ref{fig:eos_mr}) as well as for the stiffest nuclear EoS. For each case, we vary the DM parameter $G$ to obtain a band keeping the total mass above two solar masses. Since the inclusion of DM only softens the EoS, we get the largest band for the stiffest case. The bands shrink as we increase the effective mass (or soften the nuclear EoS) in both cases. This plot gives the maximum possible range that could be relevant for future detections of DM admixed NSs via GWs. The approximate range spanned by $g_1$ is 200-900 Hz, and that by $g_2$ is 140-600 Hz. We conclude that there exists a degeneracy between the NS and DM admixed NS models, like in the case of other exotic matter which needs to be taken into account in future analyses.

Typical $g$-mode frequencies of NSs lie in the range of 200-600 Hz. The appearance of exotic matter in NS interior, such as hyperons and quarks, is known to raise the $g$-mode frequencies to higher values of up to 1kHz~\citep{Tran2023, Jaikumar:2021jbw, Constantinou2021, Zhao2022}. However, in these cases, the exotic particles appear beyond a threshold density, and the effect is visible only in the case of massive NSs. In the case of hyperonic NSs, we note that the $g_1$ frequencies can be higher than 400 Hz for $M \gtrsim 1.6 M_{\odot}$ and it can go above the NS frequencies of 600 Hz for $M \gtrsim 1.8 M_{\odot}$ (see Fig.7 of~\cite{Tran2023}). Similar conclusions can be drawn for the case of hadron to quark phase transition (see Fig.10 of~\cite{Jaikumar:2021jbw} and Fig.5 of~\cite{Zhao2022}). This only applies when considering the selected case studies in these works. We could obtain higher frequencies early on if the appearance of quarks/hyperons occurs at lower densities (as seen for early crossover transition in Fig.9 of~\cite{Constantinou2021}). 

In the case of DM admixed NS (see Fig.~\ref{fig:g1_contours}(a)), we find that even the inclusion of DM as considered in this model leads to high $g$-mode frequencies of up to 900 Hz. While in the case of hyperons and hadron-quark phase transition, the $g_1$ breaches the 400 Hz mark at around $M \gtrsim 1.6 M_{\odot}$, in case of DM admixed NSs, values up to 650 Hz can be obtained for the same stellar mass. Similarly, for $M \approx 1.8 M_{\odot}$, we find $g$-mode frequencies of up to 700 Hz. This is about 100 Hz higher than seen in the case of other exotic particles. Moreover, in this model, DM is present throughout the core owing to the decay of neutrons, and there is no onset density. For other exotic matter, the $g$-mode frequencies for low mass NS ($M \approx 1 M_{\odot}$) are dictated by nuclear matter due to the absence of exotic particles. This results in $\nu_{g_1}$ in the range of 200-300 Hz for such stars. We conclude that if DM is present in NSs, we can have $\nu_{g_1}$ up to 500 Hz, even in the case of low-mass stars. {\it Observing high $g$-mode frequencies for low-mass NSs can be a tell-tale signature of DM}. 

The results from Fig.~\ref{fig:Deltag_fdm} and Fig.~\ref{fig:g1_contours} can be combined to constrain the DM fraction in the future when observation for \gm~becomes feasible. We saw that $\Delta \nu_g$ solely depends on $f_{DM}$. If we obtain a measurement on $\nu_{g,obs}$ in the future, we can use the contour in Fig.~\ref{fig:g1_contours} to obtain the maximum possible $\Delta \nu_g$. Although the measured $\nu_{g,obs}$ can be used to fit a purely nuclear EoS given that $\nu_{g,obs} < 600$, we conclude from this study that there is a possible degeneracy with DM admixed NS models. This degeneracy needs to be considered while constraining the nuclear EoS. The same observation can result from a DM admixed NS with a stiffer nuclear EoS containing a fraction of DM. Note that the lowest boundary in the $\nu-M$ plane corresponds to the stiffest EoS without DM. Thus, for a given NS mass, we can estimate the maximum possible difference between the observed \gm~frequency and the one corresponding to the stiffest EoS i.e., $\Delta \nu_{g, max} = \nu_{g,obs}(M) - \nu_{g,stiffest}(M)$. This $\Delta \nu_{g, max}$ can then be used to constrain the maximum possible DM fraction in NSs using Eq.~\ref{eqn:Deltag_fdm_fit}.

\section{Conclusions}\label{sec:conclusions}

In this work, we investigate for the first time $g$-mode oscillations in a DM admixed NS. We have used the Cowling approximation for the evaluation of $g$-modes, which agree with full-GR results within a few percent error (see Appendix~\ref{sec:appendix_fullGR}). To model the nuclear sector, we adopted the RMF formalism, where nucleonic interactions are modeled via the exchange of mesons. The mesonic fields are replaced by their mean values to obtain the EoS. For the dark sector, we consider the model for self-interacting fermionic DM based on the neutron decay anomaly. In this model, neutrons decay into the dark fermion until an equilibrium between the two is reached. 

To study the $g$-modes, we evaluate the equilibrium speeds of sound from the equilibrated EoS and derive the adiabatic speed of sound contributions from the nucleons, leptons, and the dark sector. The adiabatic speed of sound for DM depends on the DM particle fraction as well as the self-interaction parameter $G$. The difference between the two speeds of sound is calculated, which is required for the evaluation of $g$-mode.

We studied the speed of sound difference for a fixed value of $G=30$ fm$^2$. On comparison, we found that the speed difference increases upon the inclusion of DM at all densities. This also results in a corresponding rise in \bv~frequency and $g$-mode frequency. The curves for $c_s^2 - c_e^2$ peak at higher densities and do not exhibit minima when DM is included.

Upon investigating the effect of $G$ alone on $g$-mode frequencies, with the nuclear parameters kept fixed, we found a systematic increase in both $g_1$ and $g_2$ frequencies for all NS masses. We also found that the increase is higher for lower values of $G$. We also found that the rise is higher for higher NS masses. Since DM fraction is approximately a function of $M/G$~\citep{Shirke2024}, we conclude that the $g$-mode frequencies of DM admixed NSs increase with increasing DM fraction. Also, the increase for $g_1$ is higher than that for $g_2$.

For the nuclear sector, we investigated the effect of the nucleon effective mass $m^*/m$, which is the most dominant parameter in dictating the NS macroscopic properties. We found that upon fixing the DM parameter $G$ and varying $m^*/m$, the rise in the $g_1$ and $g_2$ frequencies is higher in the case of the lower effective mass (corresponding to a stiffer nuclear EoS). However, the uncertainty in $\Delta \nu_g$ is much lower in this case as compared to the one coming from the uncertainty from the dark sector. 

We also studied the $g$-mode frequencies as a function of DM fraction $f_{DM}$ and effective nucleon mass $m^*/m$. We observed that the frequencies rise with both; however, the effect of $f_{DM}$ is much stronger than the effective mass. To quantify the effect, we defined $\Delta \nu_{gi}$, the rise in the $g$-mode frequencies upon inclusion of DM when all the rest of the parameters are kept fixed. We found that the rise solely depends on the DM fraction of the DM admixed NS configuration and is independent of the underlying microscopic parameters with no explicit dependence on the stellar mass. In this sense, the relation between $\Delta \nu_{gi}$ and $f_{DM}$ is EoS independent. We also find that the dependence is linear, with the slope being higher in the case of $g_1$. We find a spread around the linear relation up to $20\%$, which can be attributed to variation induced via EoS uncertainty.   

We reported the full span of the band for \gms~of DM admixed NSs in the $\nu_{g} -M$ plane. We highlight that there is a degeneracy for DM admixed NS models with NS and it is important to consider this in future analyses involving constraints from \gms. We find that \gm~frequency can go as high as 900 Hz as compared to NS \gms, which goes up to 600 Hz. Further, the band is broader than in the case of the appearance of exotic particles such as hyperons and quarks, especially in the low NS mass regime. We conclude that observation of high \gm~frequency for low mass NSs could be a signature of DM. We also proposed a way to constrain the DM fraction within NSs using future observations of $g$-modes. We show a typical comparison between the Cowling and full-GR frameworks for \gms~of DM admixed NSs in Appendix~\ref{sec:appendix_fullGR}.

As a summary, the novel features of this work include: i) first calculation of sound speeds and that of the principal and first overtone \gms~for DM admixed NSs, ii) reporting an EoS-independent relation between the change in \gm~frequency and DM fraction (connecting \gms~with a global stellar property), and iii) reporting full span of the band for \gms~in the $\nu-M$ plane. These results are relevant in the case of a binary merger scenario, as $g$-modes can be dynamically excited during the inspiral phase due to time-varying tidal fields. Although the coupling of dynamical tides is higher in the case of fundamental $f$-modes, \gms~can be detected as they can be excited much earlier, having a lower frequency. The study of the effect of DM on $f$-modes for the model considered has already been carried out~\citep{Shirke2024}. However, \gms~are sensitive to the interior composition and future observations of \gms~can provide a crucial probe for DM in NSs.

This is the first calculation of $g$-modes for DM admixed NSs, but this formalism can be extended to other nucleonic and DM models. The calculation of the $g$-mode characteristics can be performed in full general relativity, but as we illustrated in the Appendix~\ref{sec:appendix_fullGR}, the frequencies are only affected by a few percent. One may also include the effect of rotation on the excitation of oscillation modes of a star.

\section*{Acknowledgements}

The authors are grateful to Suprovo Ghosh for discussions on the analytical calculation and for providing the speeds of sound, and to Bikram Keshari Pradhan for the numerical calculations of oscillation modes in full general relativity that were used in this work for comparison. D. C. acknowledges the warm hospitality of the California State University, Long Beach, U.S.A., where this work was initiated. S.S. and D.C. acknowledge the usage of the IUCAA HPC computing facility for numerical calculations. P.J. is supported by a grant from the National Science Foundation PHY-2310003.

\section*{Data Availability}
The data can be made available upon request.
 



\bibliographystyle{mnras}
\bibliography{refs} 



\appendix

\section{Comparison with full GR}\label{sec:appendix_fullGR}
\begin{figure}
    \centering
    \includegraphics[width = \linewidth]{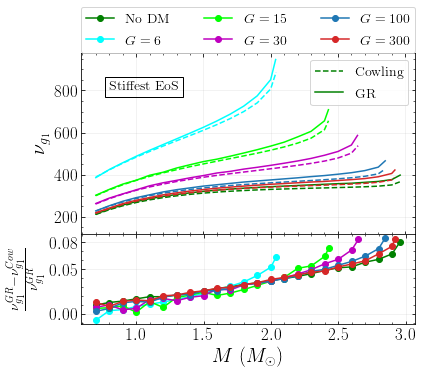}
    \caption{Comparison of $g$-mode frequency calculated using full GR and the Cowling approximation for the case without DM and with DM for different values of $G$. Here, $G$ is in fm$^2$ and $\nu_{g1}$ in Hz. The hadronic matter is described by the ``Stiffest" parametrization.} 
    \label{fig:g1_mode_comparision}
\end{figure}

We have considered Cowling approximation throughout this work to compute $g$-mode frequencies. In this section, we compare the results with the results for full-GR calculations where metric perturbations are not ignored. This is achieved by searching for low-frequency eigenvalues with the formalism developed in our previous work~\citep{Pradhan2022}. Fig.~\ref{fig:g1_mode_comparision} shows the $g_1$ mode frequency, for the ``stiffest" nuclear EoS and different select values of $G$ as a function of NS mass. Both the results from Cowling and full-GR are shown with dashed lines and solid lines, respectively. The relative error observed in the Cowling approximation is plotted in the lower panel. We observe that the Cowling approximation underestimates the $g$-mode frequency in each case. However, this error is well under 10$\%$. Hence, we restrict our work to Cowling approximation. This is consistent with the results from~\cite{Zhao2022}. 


\bsp	
\label{lastpage}
\end{document}